# DNA-based Self-Assembly of Chiral Plasmonic Nanostructures with Tailored Optical Response

Anton Kuzyk[1,*], Robert Schreiber[2,*], Zhiyuan Fan[3], Günther Pardatscher[1], Eva-Maria Roller[2], Alexander Högele[2], Friedrich C. Simmel[1], Alexander O. Govorov[3], Tim Liedl[2†]

[1]Physik-Department, Technische Universität München, Am Coulombwall 4a, 85748 Garching, Germany;
[2]Fakultät für Physik and Center for Nanoscience, Ludwig-Maximilians-Universität, Geschwister-Scholl-Platz 1, 80539 München, Germany;
[3]Department of Physics and Astronomy, Ohio University, Athens, OH, 45701, U.S.A.

*These authors contributed equally to this work
†To whom correspondence should be addressed. E-mail: tim.liedl@lmu.de

**Surface plasmon resonances generated in metallic nanostructures can be utilized to tailor electromagnetic fields[1,2]. The precise spatial arrangement of such structures can result in surprising optical properties that are not found in any naturally occurring material[3,4]. Here, the designed activity emerges from collective effects of singular components equipped with limited individual functionality. Top-down fabrication of plasmonic materials with a pre-designed optical response in the visible range by conventional lithographic methods has remained challenging due to their limited resolution, the complexity of scaling, and the difficulty to extend these techniques to three-dimensional architectures. Bottom-up molecular self-assembly provides an alternative route to create such materials which is not bound by the above limitations[5,6]. We demonstrate here how the DNA origami method[7] can be used to produce plasmonic materials with a tailored optical response at visible wavelengths. Harnessing the assembly power of 3D DNA origami[8], we arranged metal nanoparticles (NPs) with a spatial accuracy of 2 nm into nanoscale helices. The helical structures assemble in solution in a massively parallel fashion and with near quantitative yields. As a designed optical response, we generated giant circular dichroism (CD) and optical rotary dispersion (ORD) in the visible range that originates from the collective plasmon-plasmon interactions within the nanohelices. We also show that the optical response can be tuned through the visible spectrum by changing the composition of the metal nanoparticles. The observed effects are independent of the propagation direction of the incident light and can be switched by design between left- and right-handed orientation. Our work demonstrates the production of complex bulk materials from precisely designed nanoscopic assemblies and highlights the enormous potential of DNA self-assembly for the fabrication of plasmonic nanostructures. Using the presented method, strategies can be conceived to create materials of negative refractive index[9], which in turn would allow for applications such as cloaking[10] or the construction of perfect lenses[11].**

Metamaterials obtain their unique optical properties from the precise arrangement of light-sculpting elements in space. Owing to their surface plasmon resonances, metal nanostructures are particularly promising components for such applications. To obtain a metamaterial, the underlying (metal or dielectric) substrate must be structured on a length-scale smaller than the wavelength of the external electromagnetic stimulus. For the visible range this implies a characteristic scale of tens of nanometers, while for infrared (IR) wavelengths this requirement is relaxed to hundreds of nanometers. Top-down fabrication methods are readily available for the latter length scale, and therefore remarkable progress has been already achieved in the IR in the past[12-15]. Realization of materials with tailored optical response in the visible range, however, remains challenging due to the limited spatial resolution of standard lithographic techniques.

We here show that the extraordinary precision of DNA-directed molecular self-assembly makes it perfectly suited for the creation of such materials. DNA self-assembly allows for the sequence-programmable spatial arrangement of nanoscale objects, and this capability has already provided the basis for a variety of applications in structural biology, nanoelectronics, and nanophotonics[6,16-18]. An important recent development has been the introduction of the DNA origami technique, through which the assembly of nanoscale objects of unprecedented complexity has become feasible[7,8]. For DNA origami, hundreds of rationally designed "staple" oligonucleotides are hybridized to a long single-stranded DNA "scaffold" strand, which forces it to assume a specific

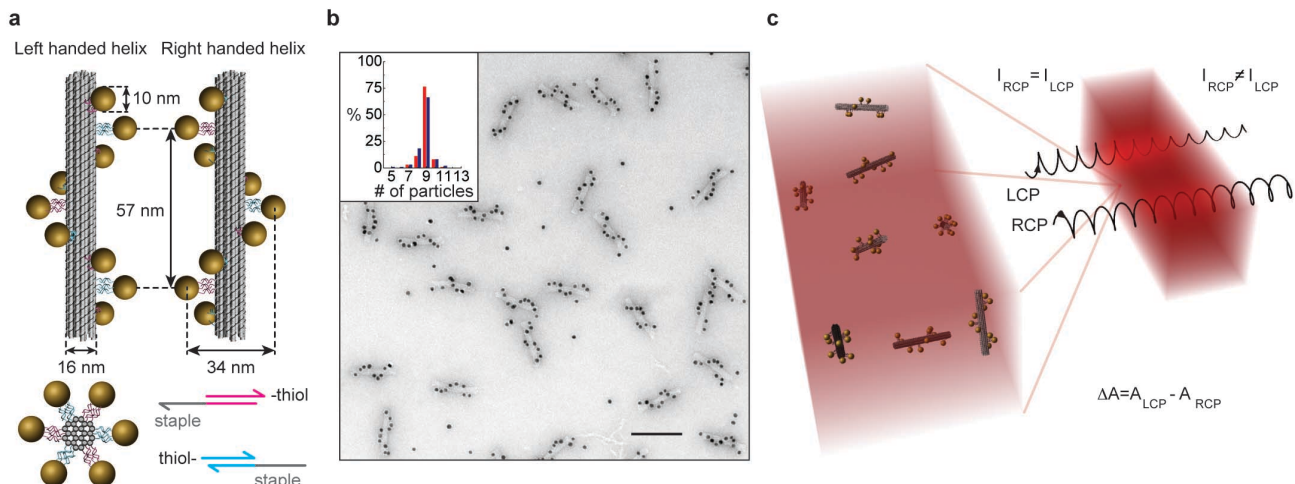

**Figure 1. Assembly of DNA origami gold particle helices and principle of CD measurements. a,** In our design, each DNA-NP assembly is composed of nine 10 nm gold particles attached in a helical arrangement to the surface of a DNA origami 24-helix bundle. Each NP attachment site consists of three 15 nt long single-stranded extensions of staple oligonucleotides. The gold nanoparticles carrying multiple thiol-modified DNA strands, which are complementary to the staple extensions, were mixed with the assembled 24-helix bundles and gel-purified. **b,** Transmission electron microscopy (TEM) image of assembled left-handed gold nanohelices (Scale bar: 100 nm). 97% of the NPs are assembled correctly. **c,** For CD measurements, the difference in absorbance (ΔA) between left-handed circularly polarized light (LCP) and right-handed circularly polarized light (RCP) of varying wavelengths through a cuvette containing a solution of gold nanohelices of a given handedness (here: left-handed) is recorded. *ΔA* is represented as CD signal in dependence of the wavelength (cf. Fig. 2).

two- or even three-dimensional shape. The resulting objects are fully addressable by their DNA sequence, and this property can be utilized to decorate origami objects with nanocomponents in a unique, sequence-specific manner.

To highlight the full power of this approach, we aimed for the creation of plasmonic structures with an optical response that (i) originates from collective effects emerging from the precise spatial arrangement of proximal nanoparticles, and that (ii) requires genuinely three-dimensional structures. Helical arrangements of metallic nanoparticles comply with these criteria. The response of helical structures to an electromagnetic field is associated with the effect of circular dichroism (CD), which denotes the differential absorption of left and right circularly polarized light. This effect is well known from "optically active" chiral molecules, and in particular from biomolecules such as DNA and proteins, which exhibit CD in the ultraviolet and IR range owing to electronic and vibronic excitations of their chiral secondary structure[19].

It has been proposed theoretically that CD can be also achieved in the visible through collective plasmon excitations in chiral assemblies of metal nanoparticles[20] and it has been shown in several experiments that the CD effect can be transferred into the visible region by combining the chiral morphology of organic molecules with the plasmonic nature of metal particles[21]. There also have been reports on DNA-directed preparation of chiral metal NP assemblies before[22-24], however, the designs used in these studies were not suited to produce controlled and appreciable optical responses.

The design of our structures is presented in Figure 1. They are based on DNA origami 24-helix bundles that offer nine helically arranged attachment sites for ssDNA-covered plasmonic particles, here gold nanoparticles with a diameter of 10 nm. The quality of the assemblies was assessed by transmission electron microscopy (TEM) (Fig 1b). Since the strength of the optical activity critically depends on the quality of the assembled structures[25], we improved our experimental protocol until we achieved a NP attachment yield of 95% and 97% per site and an overall yield of 70% and 80% of perfect assemblies of left- and right-handed helices with nine NPs, respectively (inset of Fig 1b). Most of the imperfect structures only exhibited a single defect within an otherwise well formed helix.

In simple terms, the chiral response of our helical structures to optical stimulation can be understood as follows: due to the helical geometry of the gold nanoparticles, coupled plasmon waves propagate along a helical path which leads to an increased absorption of those components of the incident light that are in accord with the handedness of the helices (Fig. 1c). Proving the success of our approach, CD spectra measured on samples with gold NP assemblies of both helicities exhibit the typical signatures expected for a chiral optical response (Figure 2a). What is more, we find excellent agreement of our experimental results with theoretical calculations based on classical electrodynamics. The strongest absorption of circularly polarized light is predicted in the vicinity of the surface plasmon frequency of the metal nanoparticles (Fig. 2b), which is perfectly reproduced by our assemblies. Strikingly, even the strength of the measured signals closely matches the magnitude predicted by the calculations.

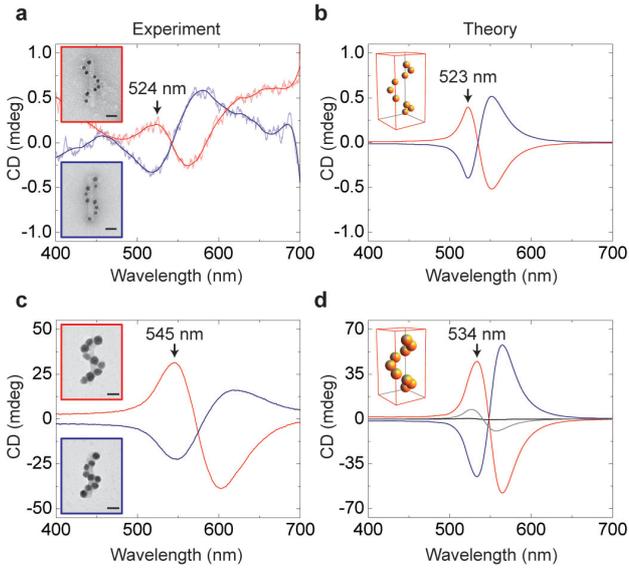

**Figure 2. Circular dichroism of self-assembled gold nanohelices. a,** A dip (peak) in the CD spectrum of right- (left-) handed 10 nm gold particle helices becomes apparent at 524 nm. This corresponds to the measured plasmon absorption peak of the samples (data not shown). Color frames indicate the correspondence between the spectra and the nanohelices presented in the TEM images (scale bars: 20 nm). **b,** The theoretically predicted CD for such geometry of 10 nm gold nanoparticles exhibits the same features as the recorded data. **c,** The optical activity of the nanohelices increases dramatically and the CD peak shifts to longer wavelengths with an increase of particle size (cf. text).

From our dipole theory[20] we expected that the CD signal becomes stronger when the particles are either larger or arranged at a closer spacing:

$$CD_{plasmon} \sim \frac{a_{NP}^{12}}{R_0^8} \quad \text{(Eq.1)},$$

where $a_{NP}$ and $R_0$ are the particle and helix radii, respectively. We therefore also tested the influence of particle size on the CD signal by employing electroless deposition of metal from solution[26-28] onto 10 nm seeding particles that were already assembled into the helical geometry described in Fig. 1. CD measurements of these "enhanced" samples showed two notable effects (Fig. 2c): i) the signal strength increased up to 400 fold for NP diameters of 16 ± 2 nm, and ii) the absorption as well as the CD peak shifted to longer wavelengths. These results are consistent with theoretical predictions for the plasmonic CD effect: A simple estimate for the increase of the CD signal based on Eq. 1 already gives the right order of magnitude assuming 16 nm particle size: (16 nm/10 nm)[12] ≈ 280, while quantitative numerical calculations predict an enhancement of ≈ 500. As in the experiments, the calculated CD for larger particles and the same helix radius is also red-shifted (Fig. 2d), which is typical for strongly interacting plasmonic nanocrystals[29-31].

The bisignate appearance of the CD signals can be understood by considering the isotropic nature of our samples, in which the nanohelices are randomly oriented with respect to the direction of the incident light. The interactions between the individual gold particles within each helix create a splitting between the longitudinal and transverse modes of the electromagnetic wave, and these modes typically have opposite chirality. As a result, the plasmonic CD spectrum acquires the characteristic dip-peak shape.

Our assembly method can be easily modified to generate a CD response also at other wavelengths. With the intention to shift the optical response of the origami nanohelices, we plated silver onto pre-assembled 10 nm gold particle helices, resulting in a silver shell of ≈ 3 nm around each of the gold NPs. Plasmon resonances in silver occur at a shorter wavelength than in gold, and as a consequence the recorded CD spectra were shifted into the blue spectral region (Fig. 3a). To fine-tune the response to intermediate wavelengths, we further employed mixtures of gold and silver ions in the plating solution to achieve the growth of alloy shells around the seed particles. Fig. 3a also contains a TEM image and a corresponding CD curve of an alloy-coated structure. As expected, the CD signal for the alloy helices is centered around an intermediate wavelength. Mixing of solutions containing different types of structures resulted in a linear superposition of the corresponding spectra (Fig 3b). These experiments clearly indicate the potential of our assembly method for the design and fabrication of optically active materials with customized spectral response.

Finally, we set out to visualize the collective optical activity of our self-assembled nanohelices in a "macroscopic" optical experiment. By exposing gold NP helices to excessive amounts of silver ions during electroless metal deposition, we achieved a giant molar CD of ≈ $10^8$ M$^{-1}$ cm$^{-1}$ (cf. 5-20 M$^{-1}$ cm$^{-1}$ per bp of dsDNA) (Fig. 3c). Droplets of such samples containing left-handed or right-handed helices were deposited next to each other on a glass slide between two crossed linear polarizers. As shown in Fig. 4, the optically active fluids rotated the polarization of linearly polarized light depending on the wavelength of the incident light and the handedness of the metal helices. This effect is closely related to the CD effect and is known as "optical rotary dispersion" (ORD). In the case of our "metafluids", polarization of red light is rotated clockwise, whereas shorter wavelengths are rotated counter-clockwise. As expected, a control sample containing isotropically dispersed 10 nm gold nanoparticles exhibited no such optical activity.

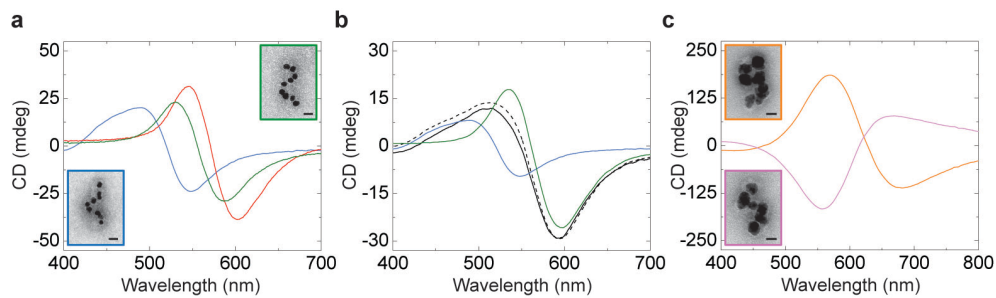

**Figure 3. Spectral tunability of optical activity. a,** The plasmon-frequency-dominated CD response of the nanohelices can be shifted to shorter wavelengths by the growth of silver shells of several nanometer thickness onto the 10 nm gold nanoparticles (blue curve). CD response at Intermediate wavelengths can be achieved by the growth of Au-Ag alloy shells (green curve). (scale bars: 20 nm). **b,** Mixing of solutions containing nanohelices of various metal compositions in one tube results in the superposition of the individual optical activities (solid black line: experiment; dashed black line: predicted superposition). The signal of the mixed fluid is rescaled to account for dilution effects due to the increase of the combined volume. **c,** If the metal deposition on the nanohelices leads to merging of the individual particles the signal strength can be increased even further (orange: left-handed helix, purple: right-handed helix). This fluid contains silver-coated nanohelices at an estimated concentration of ~ 50 pM and was analyzed in a cuvette with 3 mm optical pathlength. The molar CD is therefore on the order of $10^8$ / M · cm.

We have demonstrated that the power of the DNA origami method allows the creation of artificial plasmonic materials with a designed and tunable optical response, i.e., we achieved optical activity by self-assembly. Plasmonic circular dichroism and optical rotary dispersion were generated by arranging gold nanoparticles into helices of precisely specified geometry. The optical response could be freely tuned over a large fraction of the visible spectral range by employing different compositions of metals deposited on the 10 nm gold particles by electroless growth. Our experimental observations can be quantitatively explained by considering plasmon-plasmon Coulomb interactions within a nanoparticle helix. Since the materials produced and investigated in this study were self-assembled in solution, the process is easily scalable to larger volumes as is required for macroscopic applications based on materials with isotropic optical activity. The potential for the precise spatial arrangement of nanoparticles that possess specific electric or magnetic properties through DNA origami is immense, pointing towards the realization of a broad class of novel materials with artificial electromagnetic properties.

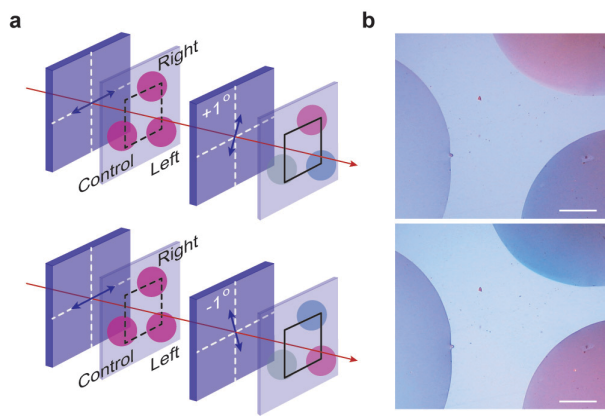

**Figure 4. Optical rotary dispersion of self-assembled gold nanohelices. a,** The macroscopic activity of the metafluid can be demonstrated by the deposition of droplets of left-handed and right-handed nanohelices between two crossed linear polarizers. **b,** Clockwise (top) or counterclockwise (bottom) rotation of the polarizers reveals the wavelength-dependent optical rotation of light transmitted through these droplets (Optical Rotary Dispersion – ORD). Right-handed helices rotate red light in the positive direction and blue light in the negative direction. Scale bar: 1 mm.

**Materials and Methods:** Experimental protocols and details on calculations and theory are available from the authors.

**Acknowledgements:** We thank Hendrik Dietz and Guillermo Acuna for experimental advice and Joachim O. Rädler and Jörg. P. Kotthaus for helpful discussions. We gratefully acknowledge Johannes Buchner and Mathias Rief for giving us access to their CD spectrometers, Eva Herold for assistance with the CD measurements, and Thomas Martin for help with the TEM. We also would like to thank David M. Smith for carefully reading the manuscript. This work was funded by the DFG Cluster of Excellence NIM (Nanoinitiative Munich) and the Volkswagen Foundation.